\documentclass{pasa}

\usepackage{graphicx}

\def\ks{km s$^{-1}$}

\def\s{$^{\prime\prime}$}

\def\cm3{cm$^{-3}$}

\def\2{$^{12}$CO}
\def\3{$^{13}$CO}
\def\8{C$^{18}$O}
\def\x{X$^{13/18}$}

\def\cm2{cm$^{-2}$}
\def\x{$X^{13/18}$}
\def\R{$R^{13/18}$}

\title[$^{13}$CO/C$^{18}$O abundance ratio in IRDC 34.43$+$0.24]{Study of the $^{13}$CO/C$^{18}$O abundance ratio towards the filamentary infrared dark cloud IRDC 34.43$+$0.24}

\author[Areal M. B. et al.]{Areal M. B.$^1$, Paron S.$^{1,2}$, Ortega M. E.$^1$ and Duvidovich L.$^{1}$
\affil{$^1$CONICET-Universidad de Buenos Aires. Instituto de Astronom\'{\i}a y F\'{\i}sica del Espacio
             CC 67, Suc. 28, 1428 Buenos Aires, Argentina}%
\affil{$^2$Universidad de Buenos Aires. Facultad de Arquitectura, Dise\~{n}o y Urbanismo. Buenos Aires, Argentina}
}%

\jid{PASA}
\doi{10.1017/pas.\the\year.xxx}
\jyear{\the\year}

\usepackage{aas_macros}
\usepackage{hyperref} 
\hypersetup{colorlinks,citecolor=blue,linkcolor=blue,urlcolor=blue}

\hypersetup{draft}

\begin{document}

\begin{frontmatter}
\maketitle

\begin{abstract}
Nowadays there are several observational studies about the \3/\8 abundance ratio (\x) towards nearby molecular clouds.
These works give observational support to the \8 selective photodissociation due to the interaction between the far ultraviolet (FUV) radiation and
the molecular gas. It is necessary to increase the sample of molecular clouds located at different distances and affected in different ways
by nearby or embedded HII regions and OB associations to study the selective photodissociation.
Using \2, \3, and \8 J=1--0 data obtained from the FOREST unbiased Galactic plane imaging survey performed with the 
Nobeyama 45 m telescope, we analyze the filamentary infrared dark cloud IRDC 34.43$+$0.24 located
at the distance of about 3.9 kpc. This IRDC is related to several HII regions and young stellar objects. 
Assuming local thermodynamic equilibrium we obtain: $0.8 \times 10^{16} <$ N(\3) $<4 \times 10^{17}$ cm$^{-2}$
(average value $= 4.2 \times 10^{16}$ cm$^{-2}$), $0.6 \times 10^{15} <$ N(\8) $<4.4 \times 10^{16}$ cm$^{-2}$  
(average value  $= 5.0 \times 10^{15}$ cm$^{-2}$), and 3 $<$ \x~$<$ 30 (average $= 8$) across the whole IRDC.
Larger values of \x~were found towards portions of the cloud related to the HII regions associated with the N61 and N62 bubbles 
and with the photodissociation regions (PDRs), precisely the regions in which FUV photons are strongly interacting with the molecular gas. 
Our result represents an observational 
support to the \8 selectively photodissociation phenomenon occurring in a quite distant filamentary IRDC.
Additionally, based on IR data from the Hi-GAL survey, the FUV radiation field was estimated in Habing units, and the 
dust temperature (T$_{d}$) and H$_{2}$ column density (N(H$_{2}$)) distribution was studied. 
Using the average of N(H$_{2}$), values in close agreement with the `canonical' abundance 
ratios [H$_{2}$]/[\3] and [H$_{2}$]/[\8] were derived. However, the obtained ranges in the abundance ratios show that if an accurate
analysis of the molecular gas is required, the use of the `canonical' values may introduce some bias. Thus, it is important 
to consider how the gas is irradiated by the far ultraviolet photons across the molecular cloud. 
The analysis of \x~is a good tool to perform that. Effects of beam dilution and clumpiness were studied.

\end{abstract}

\end{frontmatter}

\section{INTRODUCTION }
\label{sec:intro}

The study of molecular abundances towards molecular clouds located at different environments in the interstellar medium (ISM) 
is very useful for our knowledge
of the chemical evolution of the Galaxy. Taking into account that the physical processes that take place in the molecular gas 
during the born and the evolution of massive stars have a deep influence in the chemistry, the study of molecular abundances towards
massive star-forming regions is very important.

Given that the CO
is the second most abundant molecule in the ISM, its observation in millimeter wavelengths
has been crucial to probe such scenarios (e.g. \citealt{golds08,ung87,solomon87}). It is known that the far ultraviolet (FUV) radiation selectively
dissociates CO isotopes more effectively than CO (e.g. \citealt{liszt07,glass85}) thus, the analysis of the \3 and \8
emission is important for the study of the influence that the FUV photons have in the molecular gas. 
As shown by theoretical studies (see for instance \citealt{warin96} and \citealt{van88})
the selective photodissociation may result from self-shielding effects, which can depend not only on the type of isotope but also on the 
rotational quantum number J of the molecule. \citet{zie00} point out that the clumpiness of the cloud is also important in the analysis 
of the \3/\8 abundance ratio. Due to the self-shielding, the dissociation
rate increases with decreasing column density. This effect is more evident for the \8 than for the \3 as the clump size becomes smaller,  
which implies that in small clumps the \8 is almost completely dissociated and hence the \3/\8 abundance ratio becomes high.

It was found that the \8 is selectively photodissociated with respect to \3 in many relatively nearby molecular clouds that are exposed in different ways to
the radiation field \citep{yama19,kong15,minchin95a}. Moreover, this effect was analyzed in
clouds that the selective photodissociation is due   
to UV radiation from embedded OB stars (e.g. \citealt{shima14}) and clouds that are only affected by the FUV from the 
interstellar radiation field (e.g. \citealt{lin16}). Therefore, the behaviour of
the \3/\8 abundance ratio (hereafter \x) across molecular clouds could be used as a tool for indirectly evaluate 
the degree of photodissociation in the cloud \citep{paron18}.

Nowadays there are large molecular line surveys which allow us to estimate molecular abundances ratios 
towards different environments in the ISM, avoiding
the use of indirect estimations from known elemental abundances. For example, the \x~ratio usually is obtained 
from the double ratio between the $^{12}$C/$^{13}$C and $^{16}$O/$^{18}$O  \citep{wilson94}. 
Thus, increasing the sample of molecular clouds in which the \x~ratio is studied in detail from the 
direct measurements of the molecular emission is necessary.

It is known that filamentary structures are fundamental building blocks of
molecular clouds in the ISM \citep{andre14,arz11}. They contain enough
mass to give birth to high-mass stars and star
clusters (e.g. \citealt{contreras16}). These filamentary structures are usually observed as infrared dark clouds (IRDCs), 
which indeed are sites where massive stars and star clusters born \citep{rath06}. Therefore, filamentary IRDCs,
usually affected by the radiation of nearby and/or embedded HII regions, 
are interesting sources to study the behaviour of the \x~abundance ratio.

\section{Presentation of IRDC 34.43$+$0.24}

We selected the filamentary IRDC 34.43$+$0.24 \citep{xu16,sheph07} to perform an abundance ratio study. 
This filament is located at a distance of about 3.9 kpc \citep{foster14}, and the distribution and kinematic of the molecular gas 
was studied by \citet{xu16} using the \3 and \8 J=1--0 line with an angular resolution of 53\s.
As shown by these authors and \citet{sheph07}, embedded in this IRDC there are several HII regions and young stellar objects (YSOs).
In particular, the star-forming complex G34.26$+$0.15 (hereafter G34 complex) is associated with this filament, which is 
composed by several HII regions at different evolutionary stages \citep{avalos09}. As the authors point out, a cometary
ultracompact (UC) HII region (G34.26$+$0.15C) is the responsible for
the most intense emission at radio wavelengths. This UC HII region contains a cluster 
of OB stars with an O6.5 spectral type the most luminous \citep{wood89}. The region also contains
a hot core related to H$_{2}$O and OH masers (see \citealt{imai11} and
\citealt{garay85}, respectively), which indicate the existence of outflowing activity.
Bordering the most compact region of the complex there are also two large HII regions that are catalogued 
as infrared dust bubbles (N61 and N62; \citealt{church06}, HII regions G34.172+0.175 and G34.325+0.211, respectively).
Additionally, the UC HII region G34.4+0.23, which presents massive molecular outflows, lies towards the northeast of the IRDC 
\citep{sheph07}.

\begin{figure*}[tt]
\centering
\includegraphics[width=10cm]{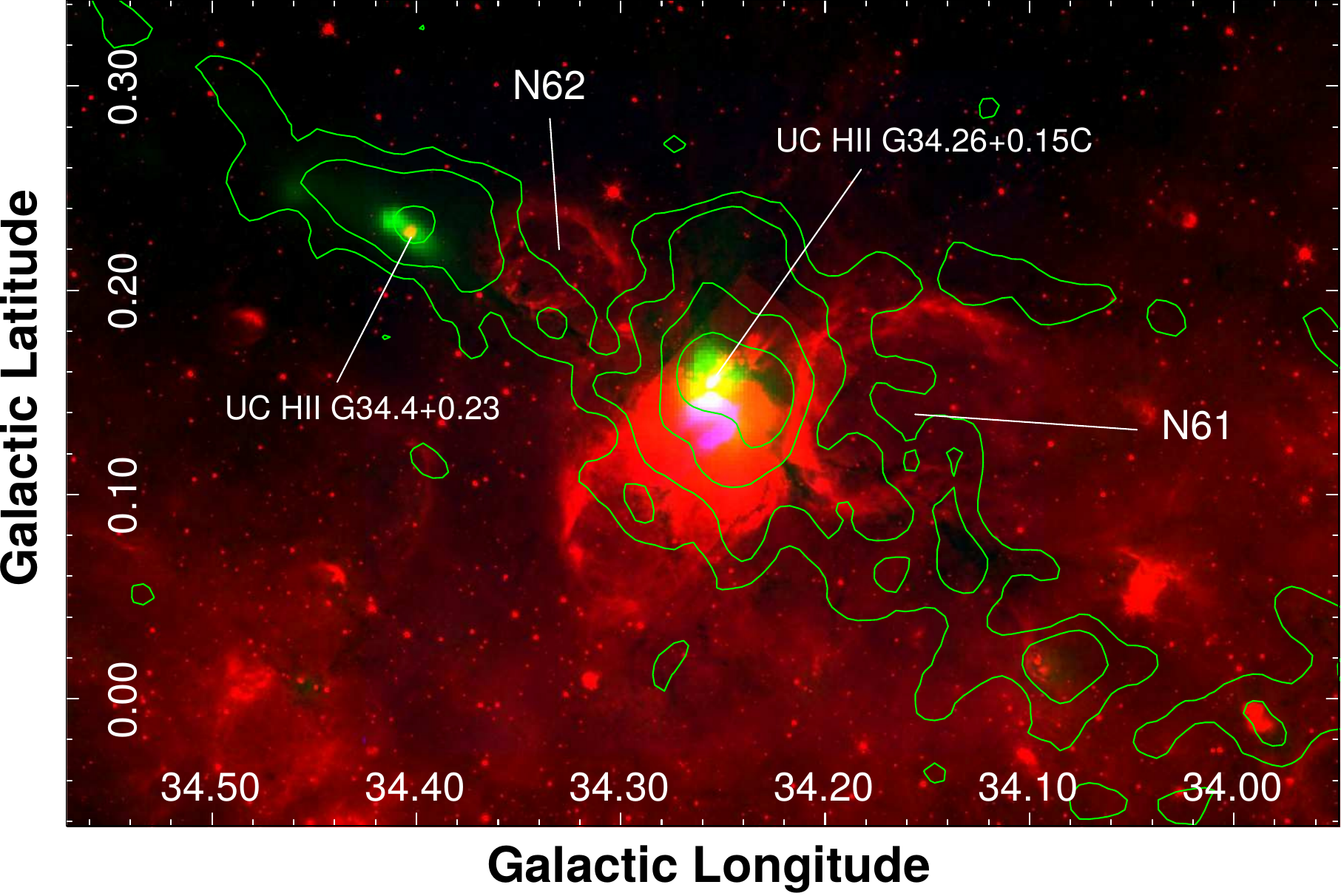}
\caption{
Three-colour image towards G34 complex displaying the {\it Spitzer}-IRAC 8 $\mu$m emission in red,
the radio continuum emission at 20 cm as extracted from the MAGPIS in blue, and the continuum emission at 1.1 mm obtained from the
Bolocam Survey in green. The green contours are the \8 J=1--0 emission as presented in Fig.\,\ref{integs} (bottom panel).}
\label{g34present}
\end{figure*}

Figure\,\ref{g34present} exhibits the IRDC 34.43$+$0.24 and G34 complex in a three-colour image, in which
the borders of the photodissociation regions (PDRs) (displayed in the {\it Spitzer}-IRAC 8 $\mu$m emission in red from the 
GLIMPSE/Spitzer survey\footnote{https://irsa.ipac.caltech.edu/data/SPITZER/GLIMPSE/}), the
distribution of the ionized gas (shown with the 20 cm emission in blue from the New GPS 20cm Survey extracted from 
the MAGPIS\footnote{https://third.ucllnl.org/gps/}), 
and the cold dust concentrations (displayed with the continuum emission at 1.1 mm in green obtained from the Bolocam Galactic Plane Survey; 
\citealt{aguirre11}) are observed.
The molecular gas distribution is shown in contours, which represent the \8 J=1--0 line integrated between 47 and 70 \ks, which is the velocity
range in which this IRDC extends (see Sect.\,\ref{data}).
As shown in Figure\,\ref{g34present}, the UC HII region G34.26$+$0.15C (G34.26 UC HII in
\citealt{sheph07}) seems to be embedded in a cold dust clump traced by the continuum emission at 1.1 mm.
Extended radio continuum emission appears southwards UC HII region G34.26$+$0.15C, which may be related to the IR-bright nebula
extending below the UC HII region as described by \citet{sheph07}.

\section{Data and the estimation of \x}
\label{data}

\subsection{Data}

The \2, \3, and \8 J=1--0 data were obtained from the FOREST unbiased Galactic plane imaging survey performed with the Nobeyama 45 m telescope
(FUGIN project; \citealt{unemoto17})\footnote{Retrieved from the JVO portal (http://jvo.nao.ac.jp/portal/) operated by ADC/NAOJ.}.
The angular resolutions are: 20\s~for the \2 data, and 21\s~for the \3 and \8 data. The spectral resolution
is 1.3 \ks~for all isotopes.

Graphical Astronomy and Image Analysis Tool (GAIA)\footnote{GAIA is a derivative of
the Skycat catalogue and image display tool, developed as part of the VLT project at ESO. Skycat and GAIA are free software under
the terms of the GNU copyright.} and tools from the Starlink software package \citep{currie14} were used to analyze the data. 
Codes in {\it python} were developed to obtain the maps of the analyzed parameters.

Additionally, we use IR data from the Hi-GAL survey \citep{molinari10}, retrieved from the 
Herschel Science Archive (HSA)\footnote{https://irsa.ipac.caltech.edu/applications/Herschel/}. In particular, maps
of Herschel-PACS at 70 and 160 $\mu$m (angular resolutions of about 6\s~and 12\s, respectively), and maps
of dust temperature and H$_{2}$ column density derived from a SED fit between 70 and 500 $\mu$m \citep{marsh17} were used (angular resolution of 36\s).

\subsection{\x~estimation}
\label{method}

\begin{figure}[h!]
\centering
\includegraphics[width=8cm]{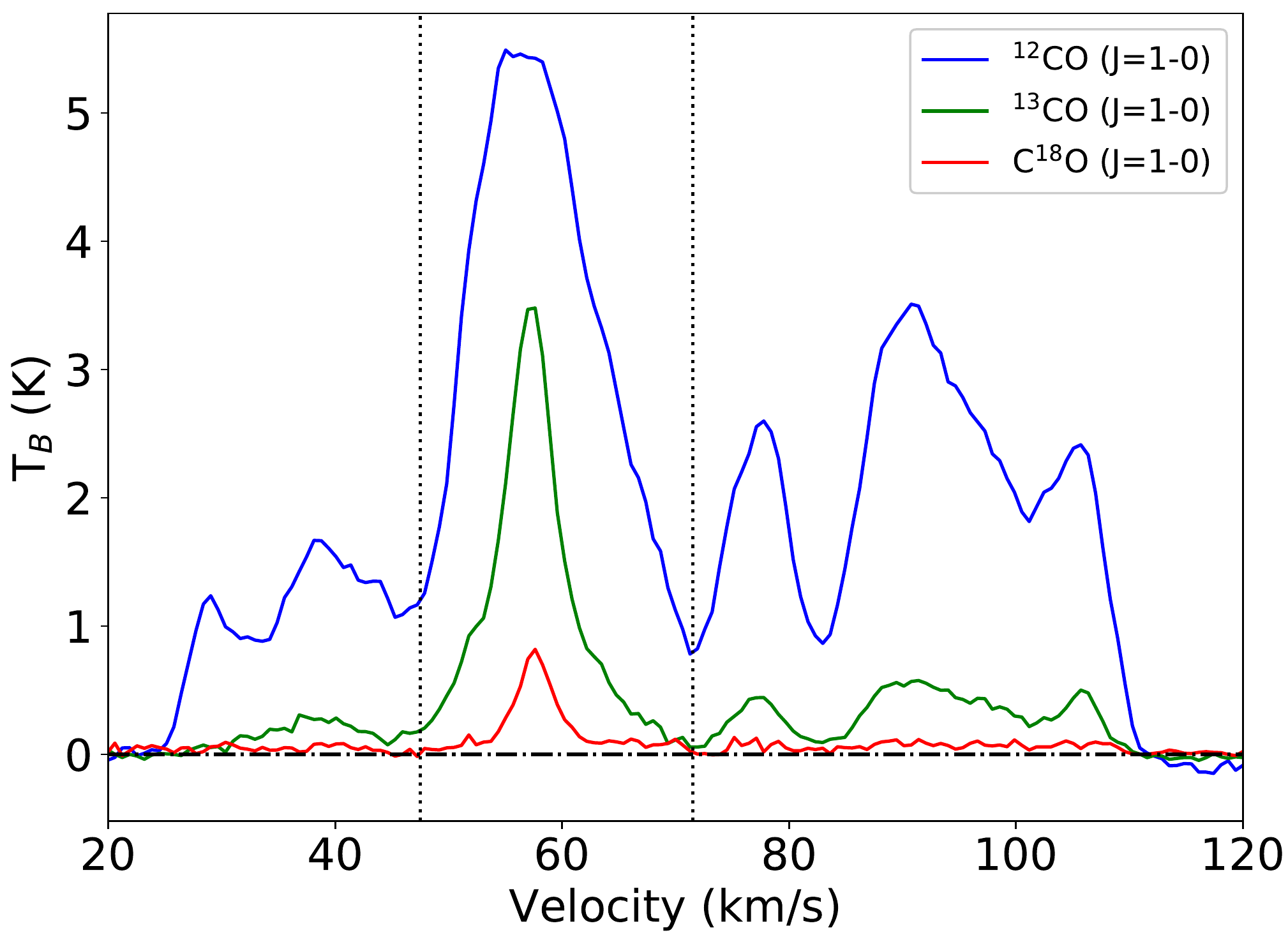}
\caption{Average \2, \3, and \8 J=1--0 spectra towards IRDC 34.43$+$0.24. The vertical dashed lines show the velocity
range in which the IRDC extends.}
\label{spectra}
\end{figure}

We carefully inspect the data cubes along the whole velocity range and find that
the systemic velocity of the \3 and \8 is quite homogeneous across the IRDC. The molecular emission extends from          
about 47 to 70 \ks. Figure\,\ref{spectra} displays average \2, \3, and \8 spectra towards the analyzed region.

In order to obtain maps of \x~(\x$=$N(\3)/N(\8)), we need to calculate the \3 and \8 column densities pixel-by-pixel.
To do that, we assume local thermodynamic equilibrium (LTE) and a beam filling factor of 1, and 
we follow the standard procedures (e.g. \citealt{mangum15}). The optical depths
($\tau_{\rm ^{13}CO}$ and $\tau_{\rm C^{18}O}$) and column densities
(N(\3) and N(\8)), were derived from the following equations:

\begin{equation}
  \tau_{\rm ^{13}CO} = - ln\left(1 - \frac{T_{mb}({\rm ^{13}CO})}{5.29\left[\frac{1}{e^{5.29/T_{ex}}-1} - 0.164\right]}\right)    
\label{tau13}
\end{equation}

\begin{equation}
{\rm N(^{13}CO)} = 2.42 \times 10^{14}~\frac{T_{ex} + 0.88}{1 - e^{\frac{-5.29}{T_{ex}}}} 
\int{\tau_{\rm ^{13}CO}{\rm dv}}
\label{N13}
\end{equation}
with
\begin{equation}
\int{\tau_{\rm ^{13}CO}{\rm dv}} = \frac{1}{J(T_{ex}) - 0.868} \frac{\tau_{\rm ^{13}CO}}{1-e^{-\tau_{\rm ^{13}CO}}}
\int{T_{\rm mb}({\rm ^{13}CO}){\rm dv}}
\label{integ13}
\end{equation}
\begin{equation}
  \tau_{\rm C^{18}O} = - ln\left(1 - \frac{T_{mb}({\rm C^{18}O})}{5.27\left[\frac{1}{e^{5.27/T_{ex}}-1} - 0.166\right]}\right)    
\label{tau18}
\end{equation}
\begin{equation}
{\rm N(C^{18}O)} = 2.42 \times 10^{14}~\frac{T_{ex} + 0.88}{1 - e^{\frac{-5.27}{T_{ex}}}} 
\int{\tau_{\rm C^{18}O}{\rm dv}}
\label{N18}
\end{equation}
with
\begin{equation}
\int{\tau_{\rm C^{18}O}{\rm dv}} = \frac{1}{J(T_{ex}) - 0.872} \frac{\tau_{\rm C^{18}O}}{1-e^{-\tau_{\rm C^{18}O}}}
\int{T_{\rm mb}({\rm C^{18}O}){\rm dv}}
\label{integ18}
\end{equation}

\noindent in which correction for high optical depths was applied \citep{frer82}. This correction is indeed required in the case of \3 but not for \8 because it is mostly optically thin.

The $J(T_{ex})$ parameter is $\frac{5.29}{exp(\frac{5.29}{T_{ex}}) - 1}$  and 
$\frac{5.27}{exp(\frac{5.27}{T_{ex}}) - 1}$ for Eqs.\,(\ref{integ13}) and\,(\ref{integ18}), respectively.
$T_{\rm mb}$ and  $T_{ex}$ are the main brightness temperature
and the excitation temperature, respectively. Assuming that the \2 J=1--0 emission is optically thick, it is possible 
to derive $T_{ex}$ from:
\begin{equation}
T_{ex} = \frac{5.53}{{\rm ln}[1 + 5.53 / (^{12}T_{\rm peak} + 0.818)]}
\label{tex}
\end{equation}

\noindent where $^{12}T_{\rm peak}$ is the \2 peak temperature. Thus from maps of $^{12}T_{\rm peak}$, 
maps of excitation temperatures ($T_{ex}$) are obtained. Before obtaining the column densities maps, 
we generate \3 and \8 optical depths ($\tau^{13}$ and $\tau^{18}$) maps and check the values. 

\begin{figure}[h!]
\centering
\includegraphics[width=8.5cm]{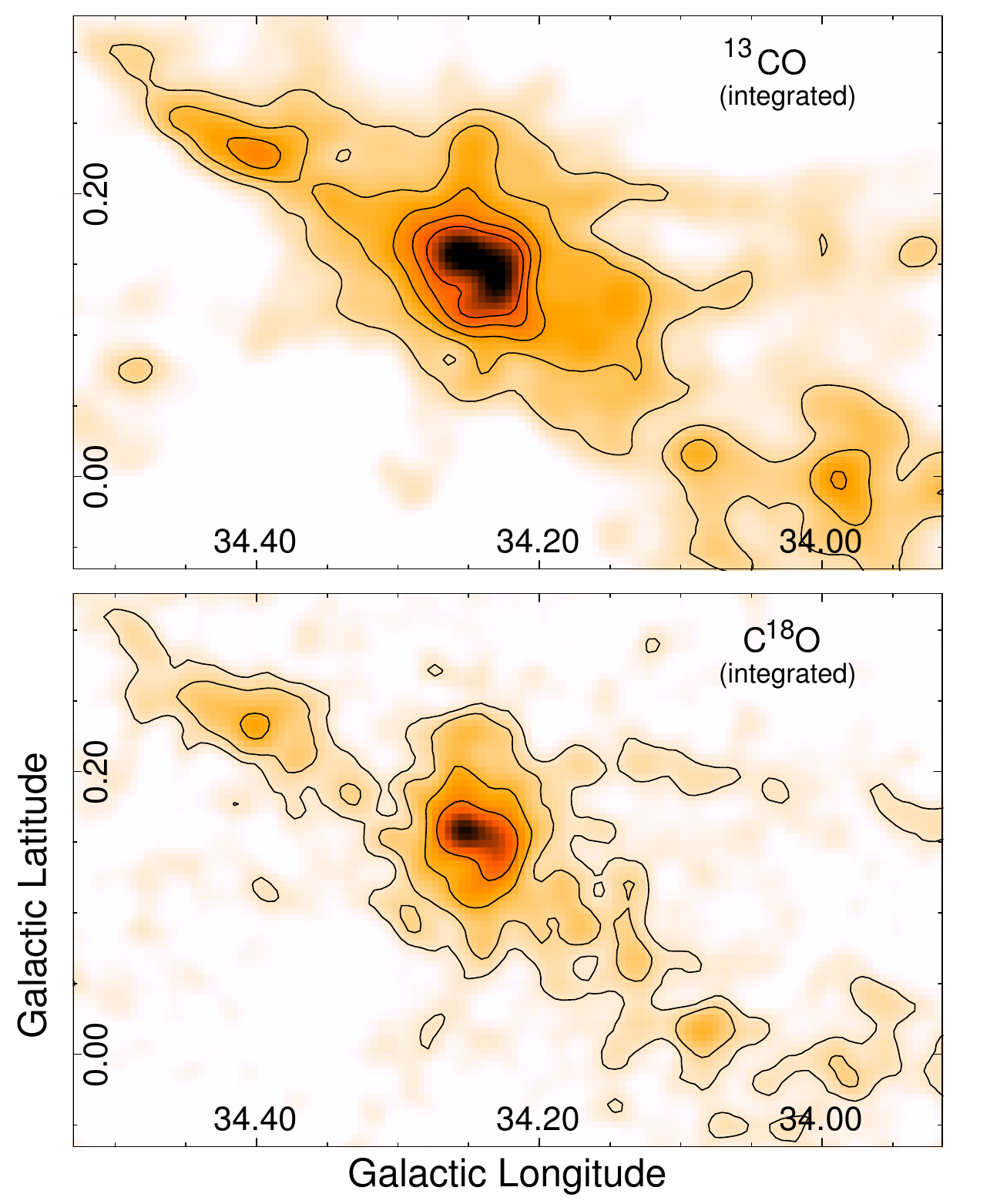}
\caption{Top and bottom panels: maps showing the \3 and \8 J=1--0 line integrated between 47 and 70 \ks, respectively.
The contours levels are 22, 30, 40, 50, and 60 K \ks, and 4, 6, 10, and 16 K \ks~for the \3 and \8, respectively. The angular resolution
is 21\s. The sigma levels of these integrated maps are: $\sigma_{13} = 4.5$ and $\sigma_{18} = 1.0$ K \ks.}
\label{integs}
\end{figure}

In some pixels, we notice that $\tau^{13}$ becomes undefined due to a negative argument in the logarithm of Eq.\,\ref{tau13}. 
It is likely that this issue is a consequence of high-saturation in the line emission. 
We investigate the spectra corresponding to 
such pixels and we find that the \2 profiles indeed present signs of saturation and self-absorption 
which may affect the estimate of T$_{ex}$. Also some \3 spectra in such pixels, and in others, mainly
at the vicinity of the UC G34.26$+$0.15C,  
present kinematic signatures of infall (as found by \citealt{xu16}) that prevent us to obtain reliable 
values of column densities. 

\begin{figure}[h!]
\centering
\includegraphics[width=8.4cm]{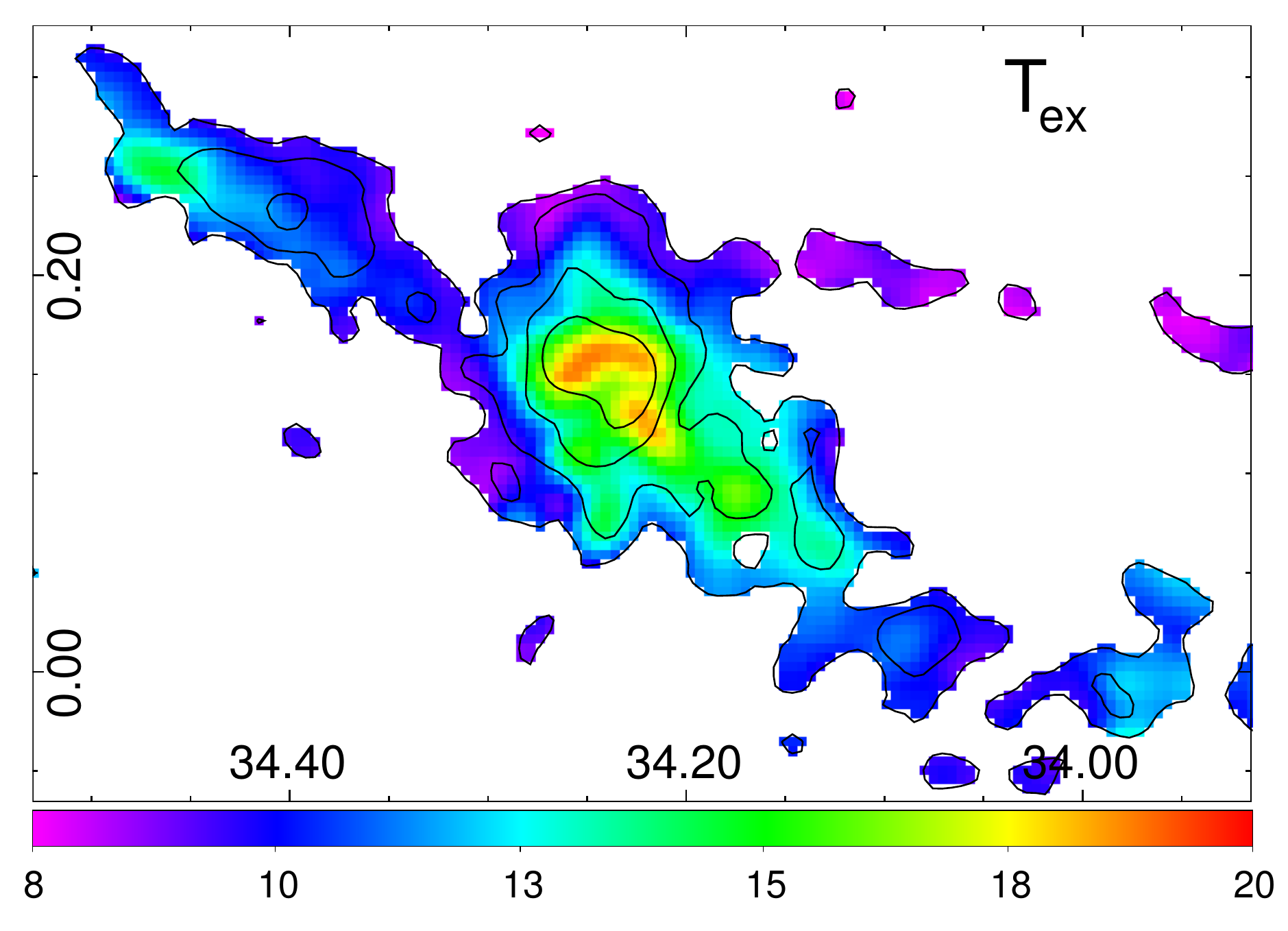}
\caption{Excitation temperature map derived from the \2 J=1--0 emission in the region of the \8 emission.
For reference, the \8 contours presented in Fig.\,\ref{integs} (bottom panel) are included.}
\label{Tex}
\end{figure}

\section{Results}

\subsection{Molecular gas}

Figure\,\ref{integs} displays the molecular gas distribution associated with the IRDC 34.43$+$0.24 in the
\3 and \8 J=1--0 line integrated between 47 and 70 \ks, with an angular resolution of 21\s. The lowest 
contour of the integrated \8 is 4$\sigma$. We consider it as a threshold to study the abundance ratio and others physical parameters derived in this work, i.e., all the analysis is done for the region contained by this \8 contour.

\begin{figure}[h!]
\centering
\includegraphics[width=8.5cm]{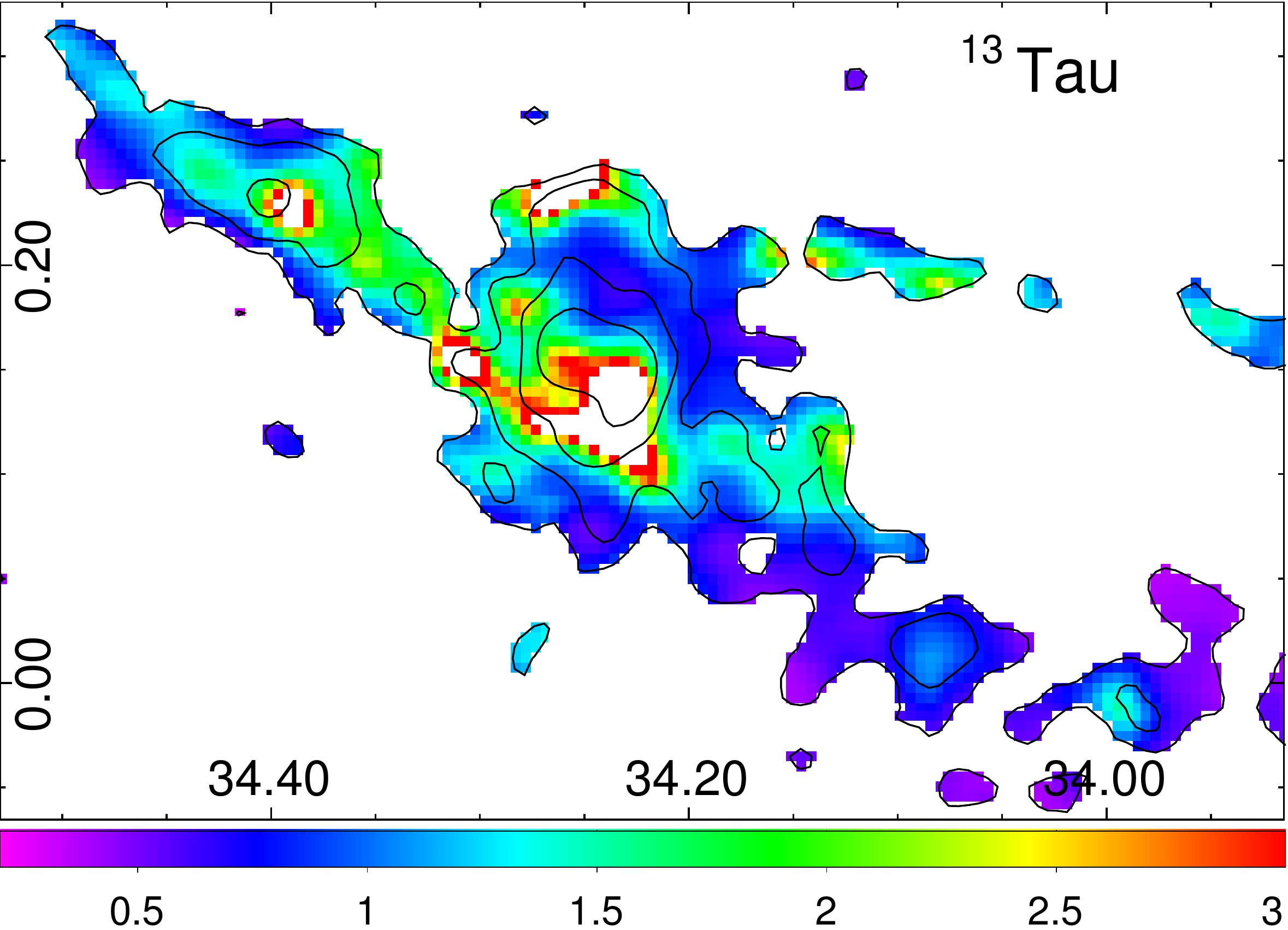}
\includegraphics[width=8.5cm]{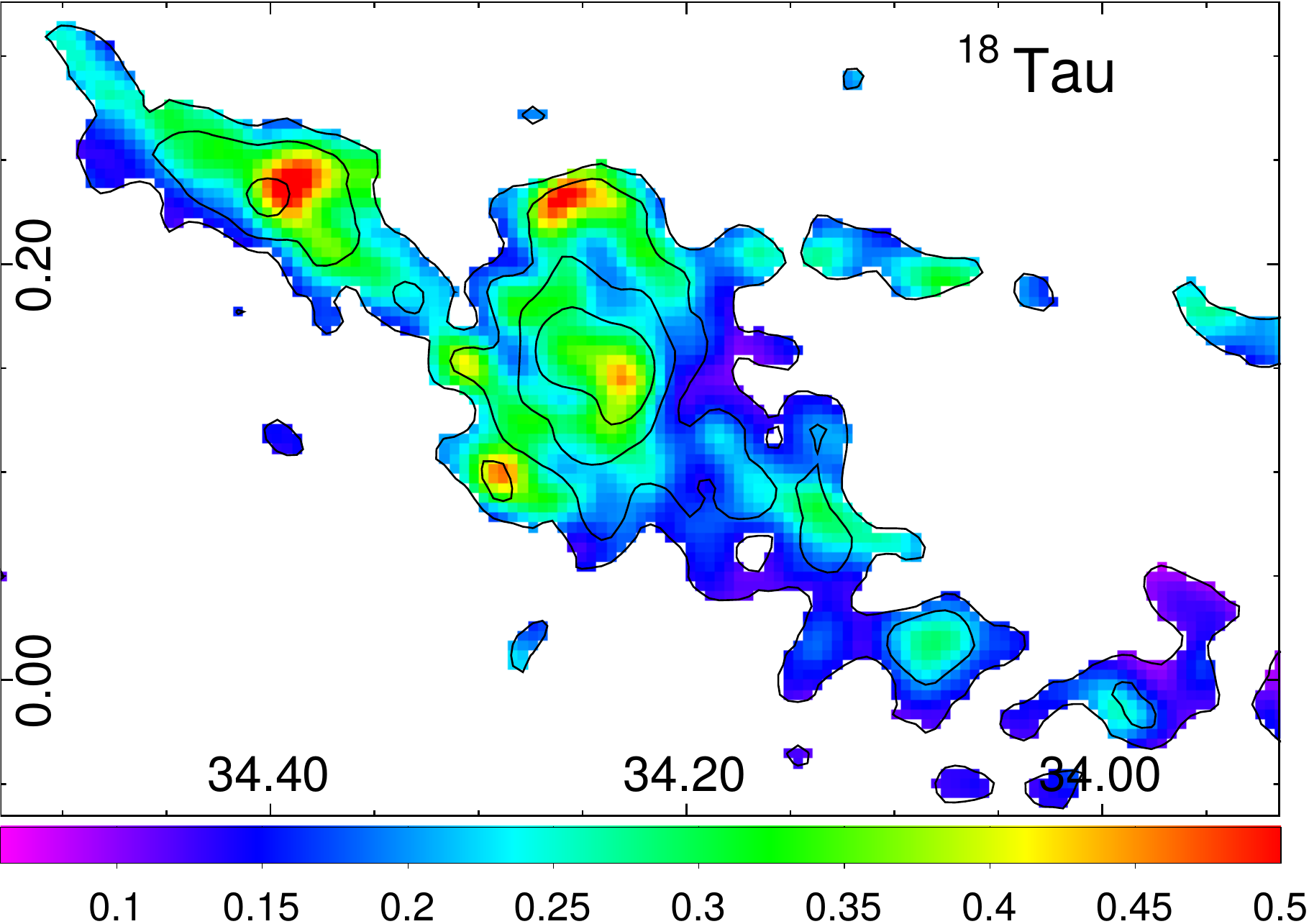}
\caption{Maps showing the \3 and \8 optical depths (top and bottom panels, respectively).
Contours of the integrated \8 J=1--0
emission are included for reference. }
\label{Taus}
\end{figure}

\begin{figure}[h!]
\centering
\includegraphics[width=8.5cm]{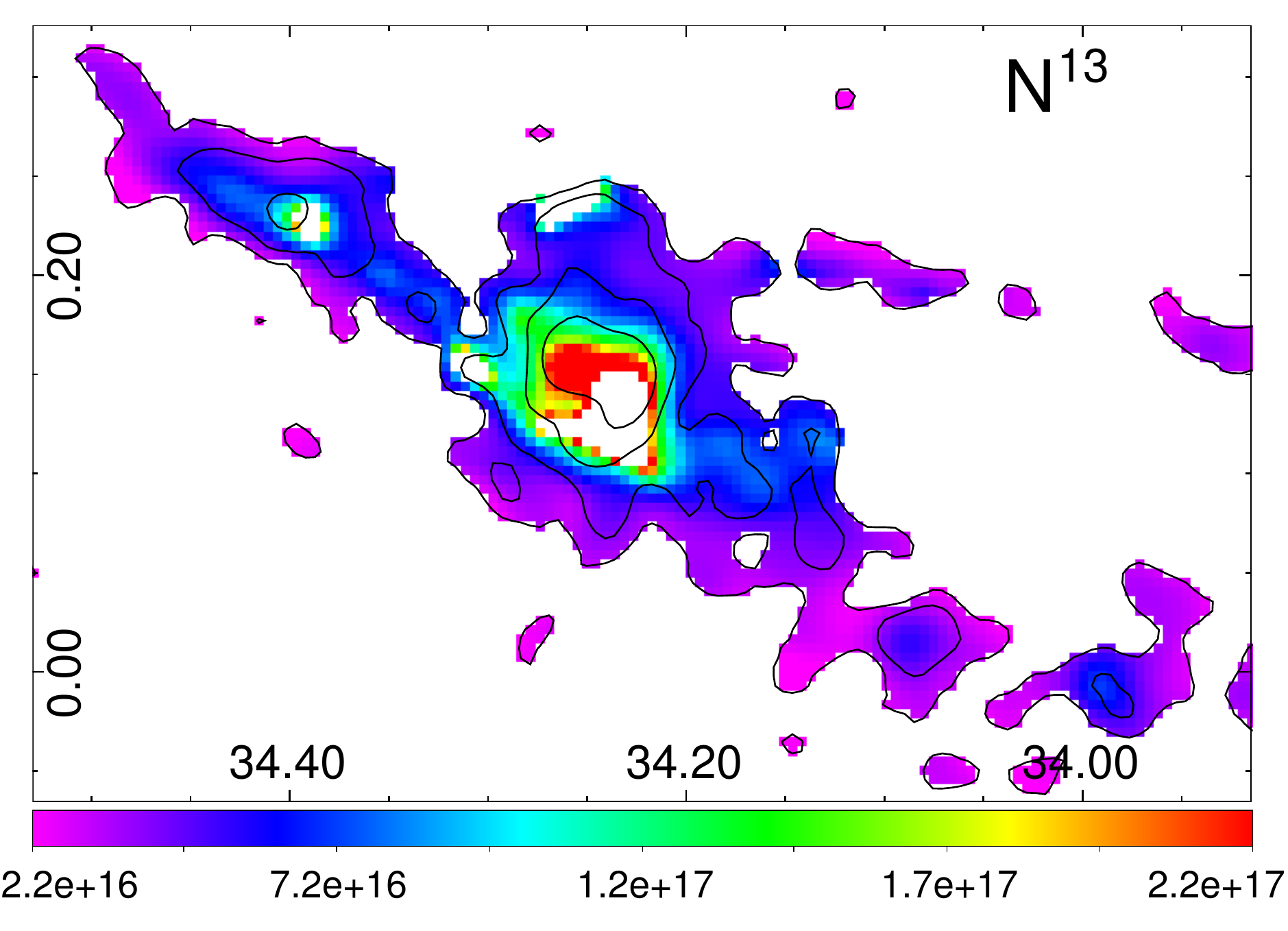}
\includegraphics[width=8.5cm]{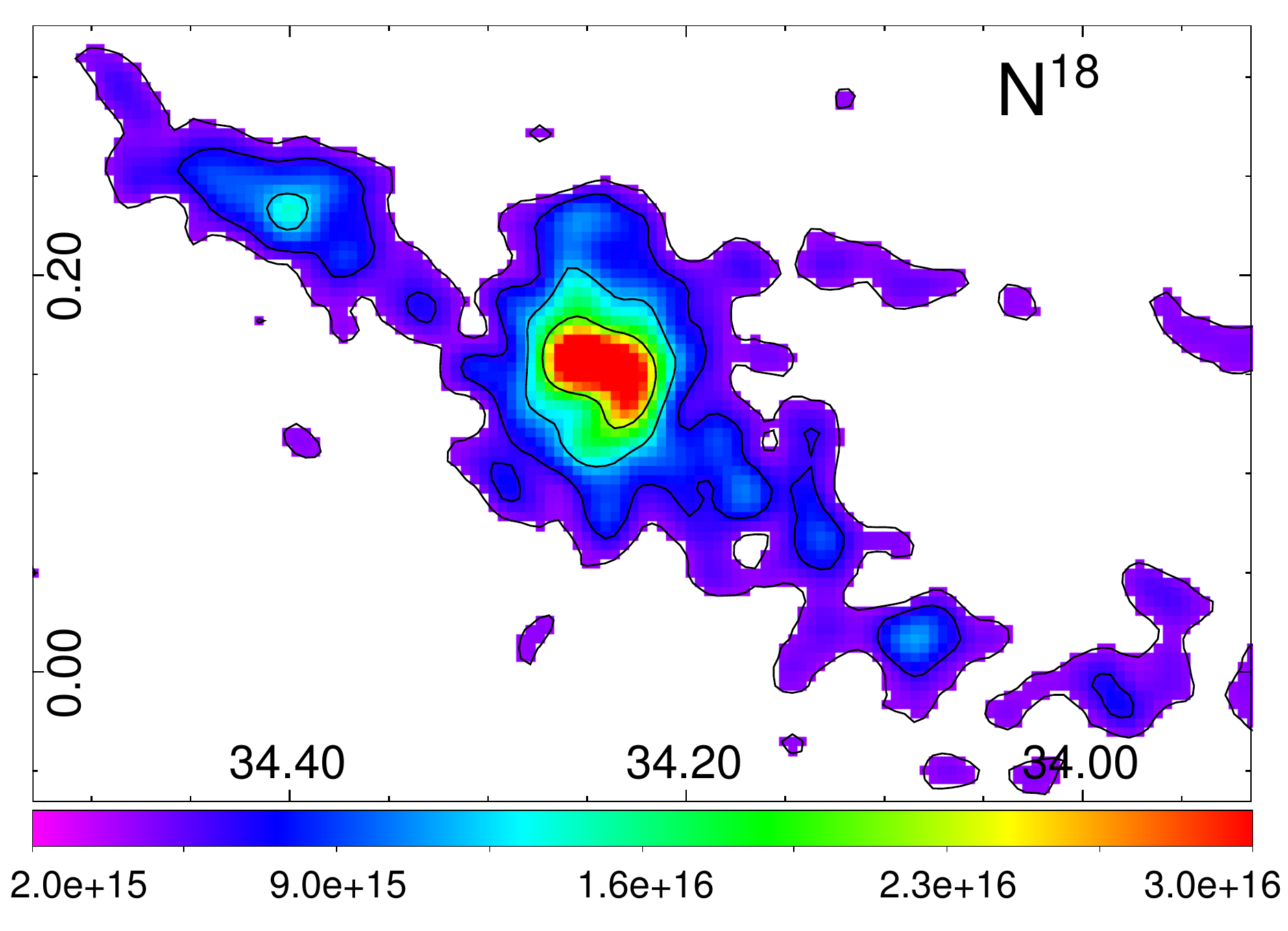}
\includegraphics[width=8.5cm]{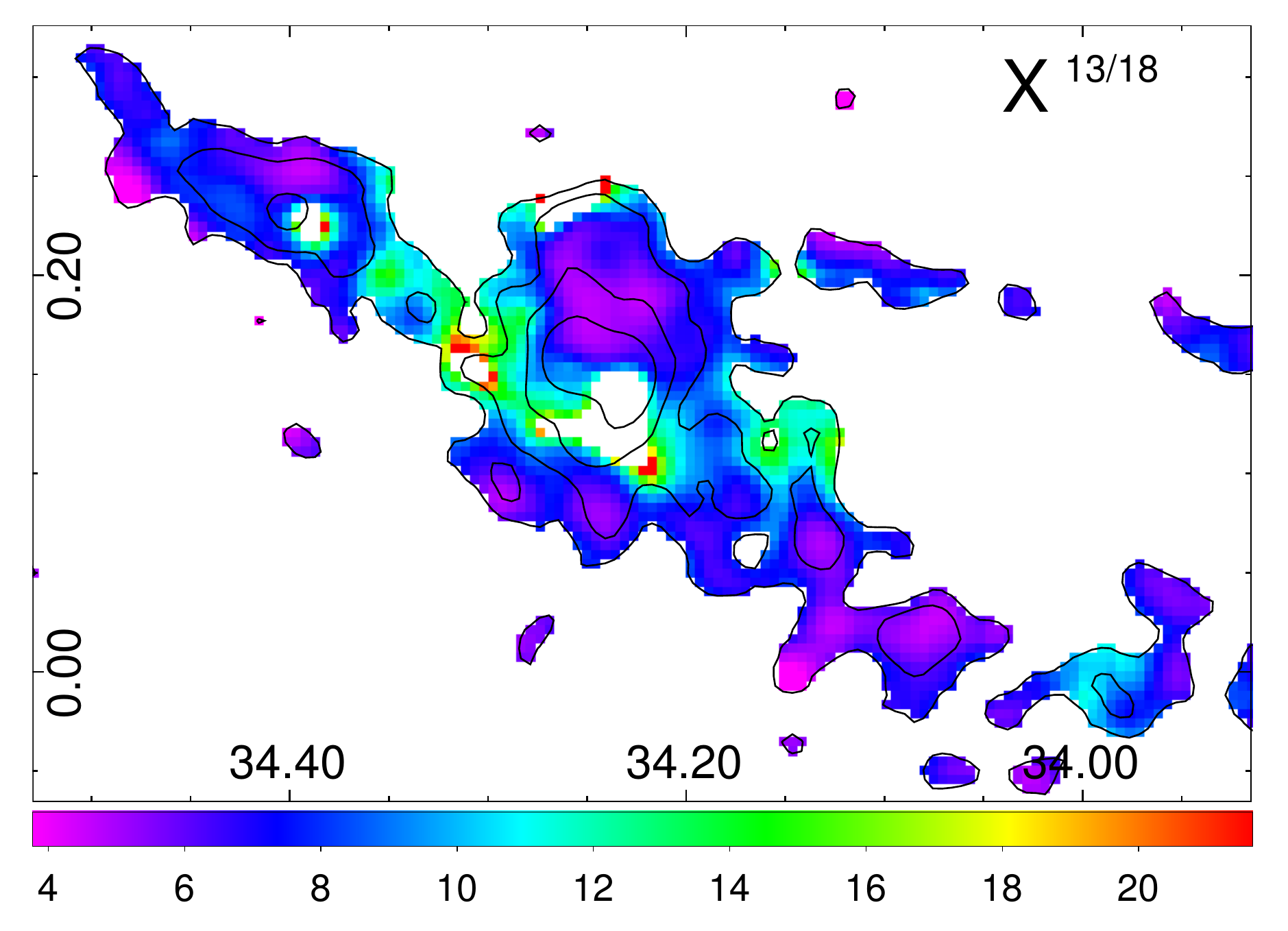}
\caption{Top and middle panels: maps showing the \3 and \8 column densities, respectively. 
The colourbars are in units of cm$^{-2}$. Bottom panel: the abundance ratio \x. Contours of the integrated \8 J=1--0 
emission are included for reference. }
\label{result10}
\end{figure}

\begin{figure}[h!]
\centering
\includegraphics[width=8.5cm]{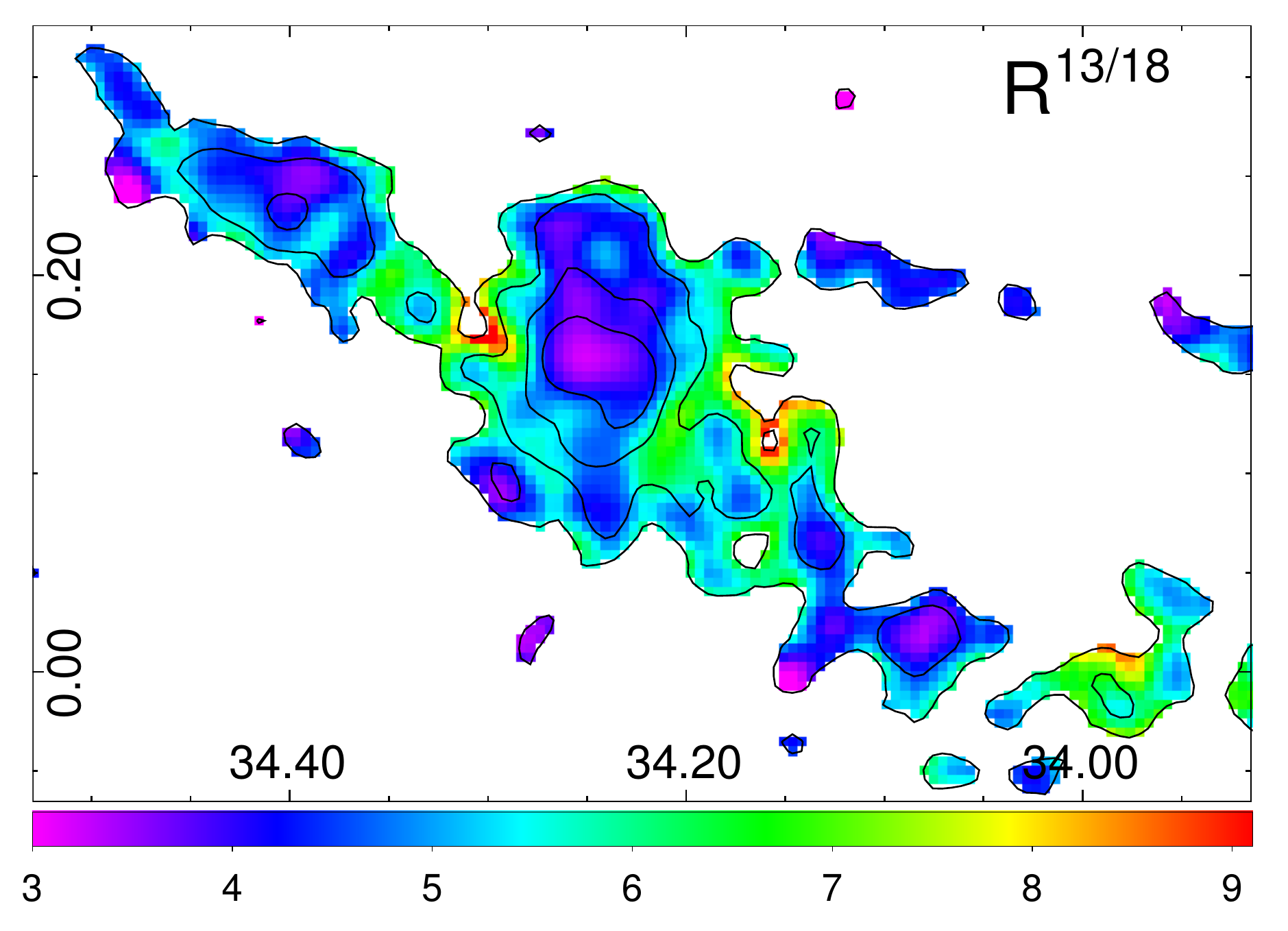}
\caption{Integrated line ratio (\R). Contours of the integrated \8 J=1--0
emission are included for reference. }
\label{ratioInt}
\end{figure}

Figure\,\ref{Tex} shows a map of the excitation temperature obtained from the \2 J=1--0 emission following Eq.\,\ref{tex}, and
Figure\,\ref{Taus} displays the maps of the obtained \3 and \8 optical depths.
Figure\,\ref{result10} presents the obtained maps of the \3 and \8 column densities and the derived abundance ratio \x. 
Figure\,\ref{ratioInt} shows the integrated line ratio (\R$=\int{\rm T^{13}dv}/\int{\rm T^{18}dv}$) and 
Table\,\ref{ns} presents ranges and averages values of the physical parameters obtained from our analysis.

\begin{table*}
\caption{Ranges of physical parameters.}
\label{ns}
\centering
\begin{tabular}{l c c c c c c c}
\hline\hline
             & T$_{ex}$ & $\tau^{13}$  & $\tau^{18}$    &    N(\3)                            &  N(\8)   &                     \x  & \R    \\
             &    [K]   &              &                &    [$\times10^{17}$ cm$^{-2}$]      &  [$\times10^{16}$ cm$^{-2}$] &       &  \\    
\hline                     
Ranges       &  8 -- 19  & 0.2 -- 5.7  &  0.06 -- 0.50 &   0.08 -- 4.00           &  0.06 -- 4.40           &  3 -- 30 & 2.4 -- 9.6 \\
Average      &   11      &    0.8      &   0.18      &      0.42                &     0.50                &     8      &  5.2   \\
\hline
\end{tabular}
\end{table*}

\begin{figure}[h!]
\centering
\includegraphics[width=8.3cm]{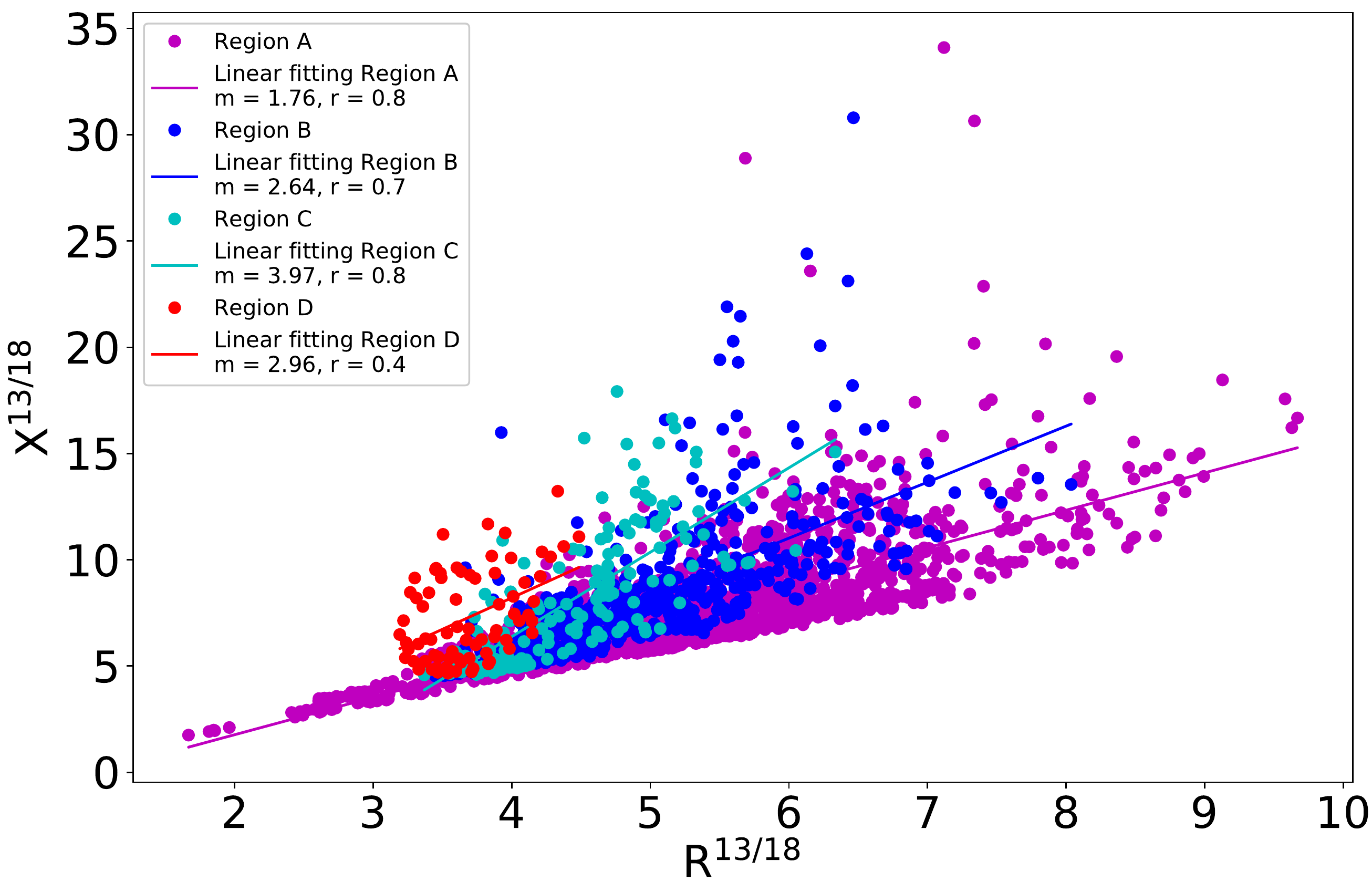}
\caption{Abundance ratio (\x) vs. integrated line ratio (\R). Region A, B, and C correspond to pixels between the \8 J=1--0 
contours 4 and 6, 6 and 10, and 10 and 16 K \ks, respectively, and Region D corresponds to pixels within the 16 K \ks~\8 contour.
Slopes (m) and correlations factors (r) from  linear fittings are included. }
\label{RvsX}
\end{figure}

Figure\,\ref{RvsX} shows the relation between the abundance ratio \x~and the integrated line ratio \R~obtained from 
different sectors of the molecular cloud defined by the 
\8 contours: Region A (pixels lying between the contours at 4 and 6 K \ks), Region B (pixels between contours 6 and 10 K \ks),
Region C (pixels between contours 10 and 16 K \ks), and Region D (pixels within the contour at 16 K \ks). The slopes (m) and 
correlations factors (r) from linear fittings for each case are included in the figure.

\subsection{Determining the FUV radiation field}

\begin{figure}[h!]
\centering
\includegraphics[width=8.5cm]{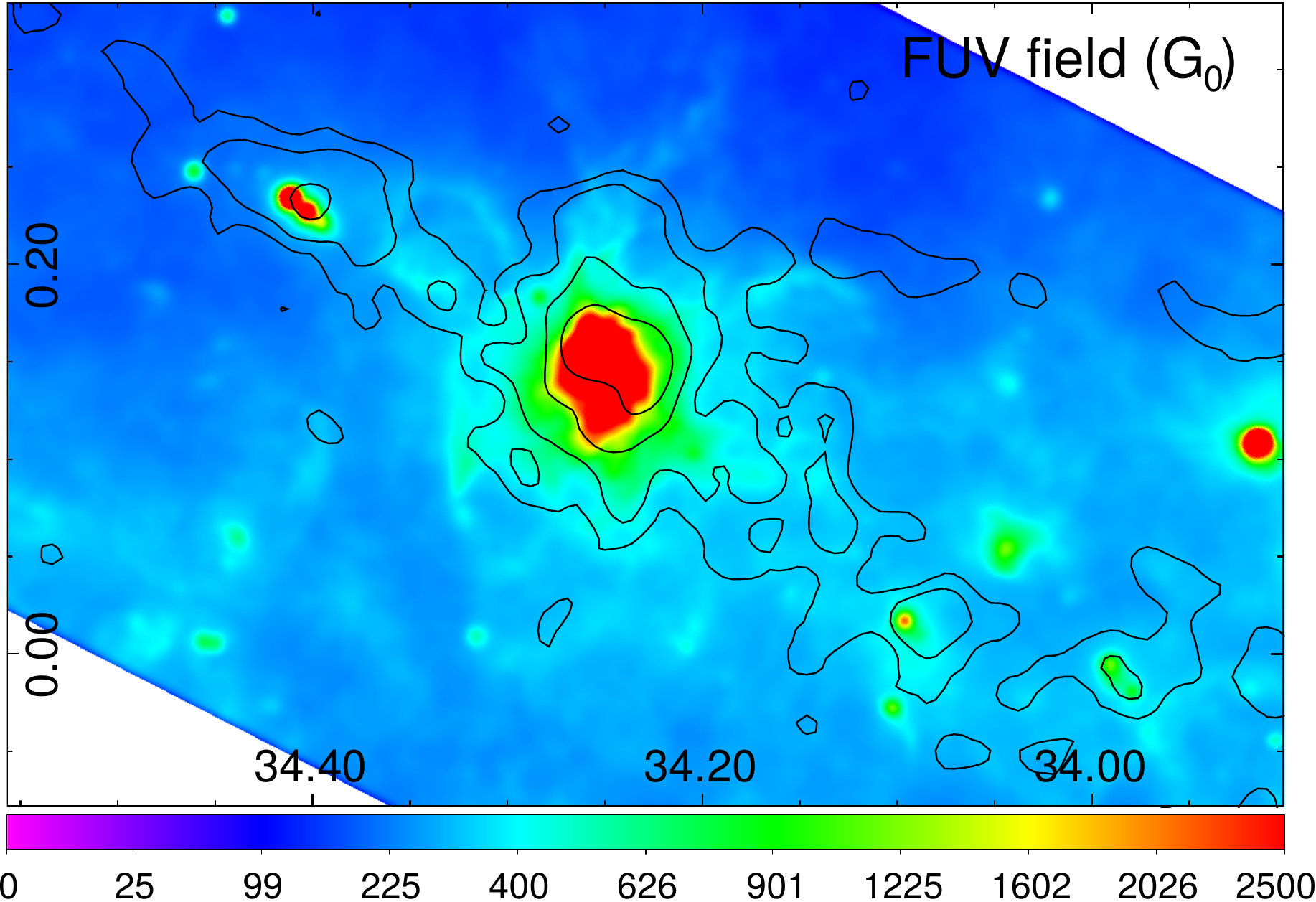}
\caption{Far-ultraviolet flux G$_{0}$ in units of Habing field. The angular resolution of the image is 21\s. 
Contours of the integrated \8 J=1--0 emission are included for reference.}
\label{uvfield}
\end{figure}

\begin{figure}[h!]
\centering
\includegraphics[width=8cm]{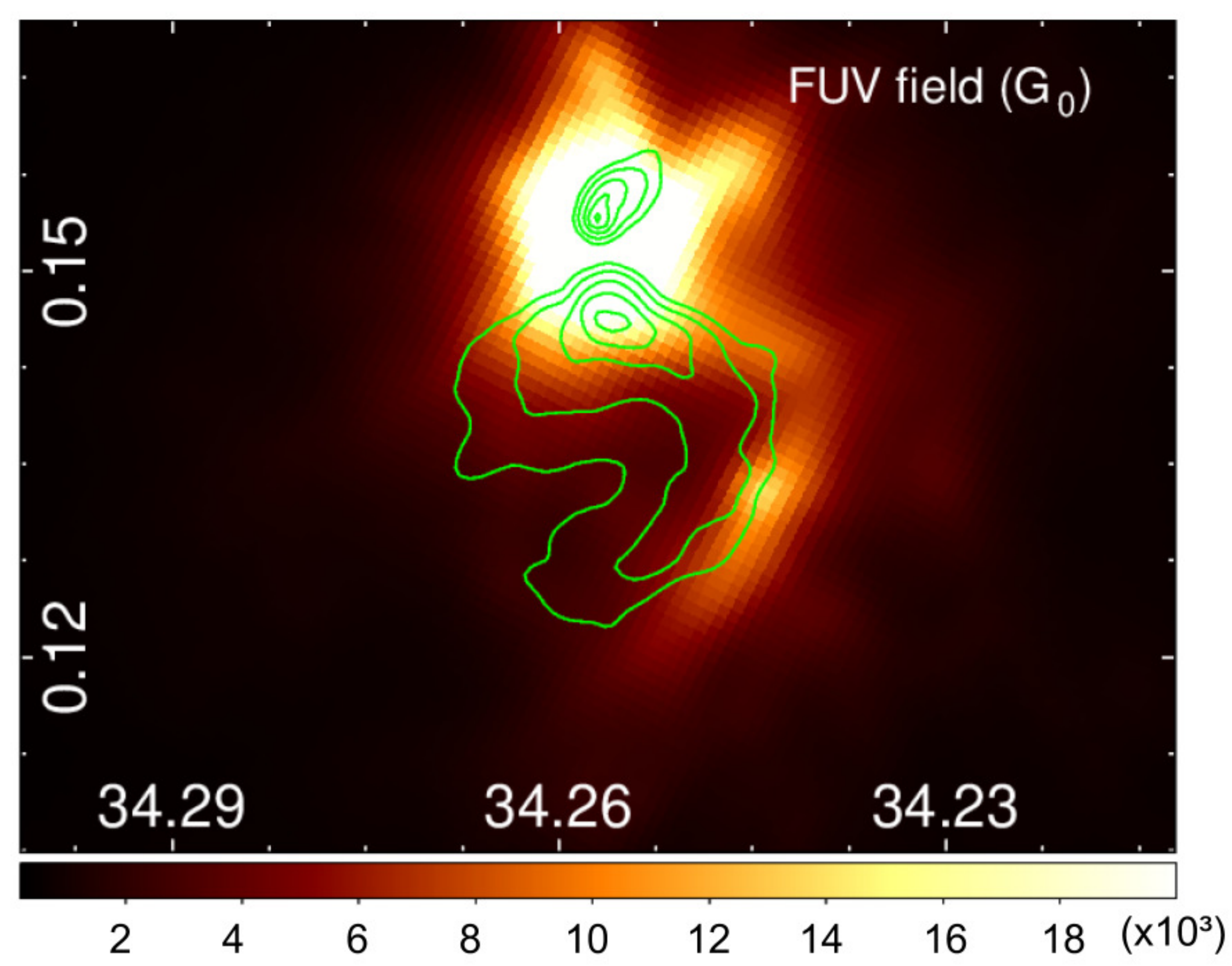}
\caption{Far-ultraviolet flux G$_{0}$ in units of Habing field towards the center of the analyzed region. 
The angular resolution of the image is about 12\s. Contours of the radio continuum emission at 20 cm are presented with 
levels of 0.02, 0.05, 0.10, 0.15, and 0.20 Jy beam$^{-1}$.}
\label{uvfieldS}
\end{figure}

Given that the clouds seen in far-IR (FIR) are mainly heated by FUV radiation, it is possible to estimate 
the FUV radiation field from the observed FIR intensity ($I_{FIR}$) (see \citealt{kramer08}).
We compute the FUV radiation field from the FIR intensity in 
the 60--200 $\mu$m range using the Herschel-PACS 70 $\mu$m and 160 $\mu$m maps. These maps
were convolved to the resolution of the molecular data, and following the procedure explained in \citet{rocca13}, we generate
the FUV radiation map in units of the Habing field from:
\begin{equation}
G_{0} = \frac{4 \pi I_{FIR}}{1.6 \times 10^{-3}~{\rm erg~cm^{-2}~s^{-1}}},
\label{uv}
\end{equation}
which is presented in Figure\,\ref{uvfield}. Besides, in Figure\,\ref{uvfieldS} we present the same FUV radiation field
towards the center of the analyzed region where the UC HII region G34.26+0.15C and the radio continuum shell lie. To do this map 
with the best angular resolution as possible, the image of PACS 70 $\mu$m was convolved to the angular resolution of the 
PACS 160 $\mu$m map (about 12\s), and the same procedure done for the map of Fig.\,\ref{uvfield} was applied. 
Contours of the radio continuum emission are included to analyze the ionized gas component in relation to the FUV radiation.

\subsection{T$_{dust}$ and N(H$_{2}$) towards the center of the IRDC}

\begin{figure}[h!]
\centering
\includegraphics[width=8.5cm]{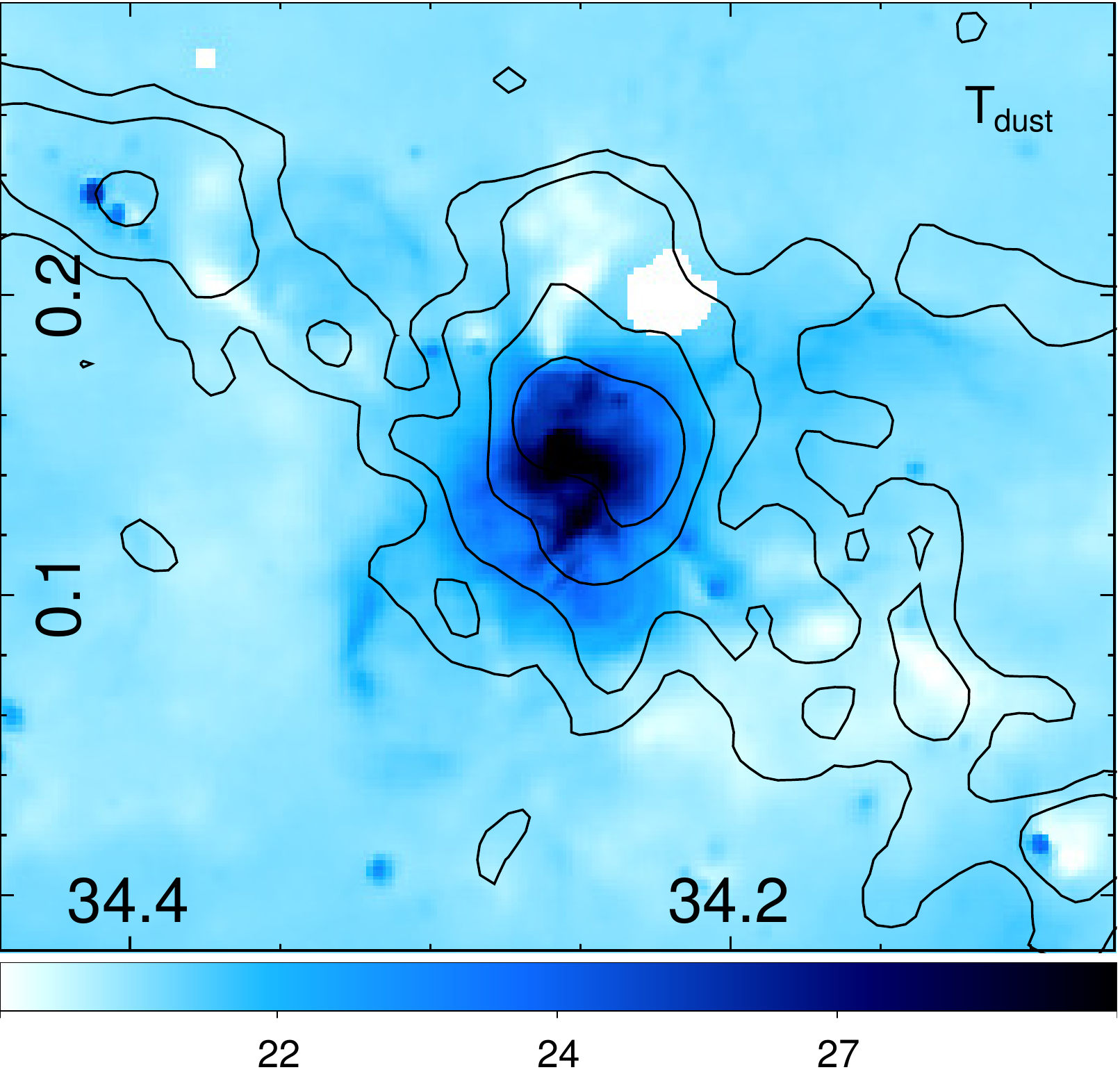}
\includegraphics[width=8.5cm]{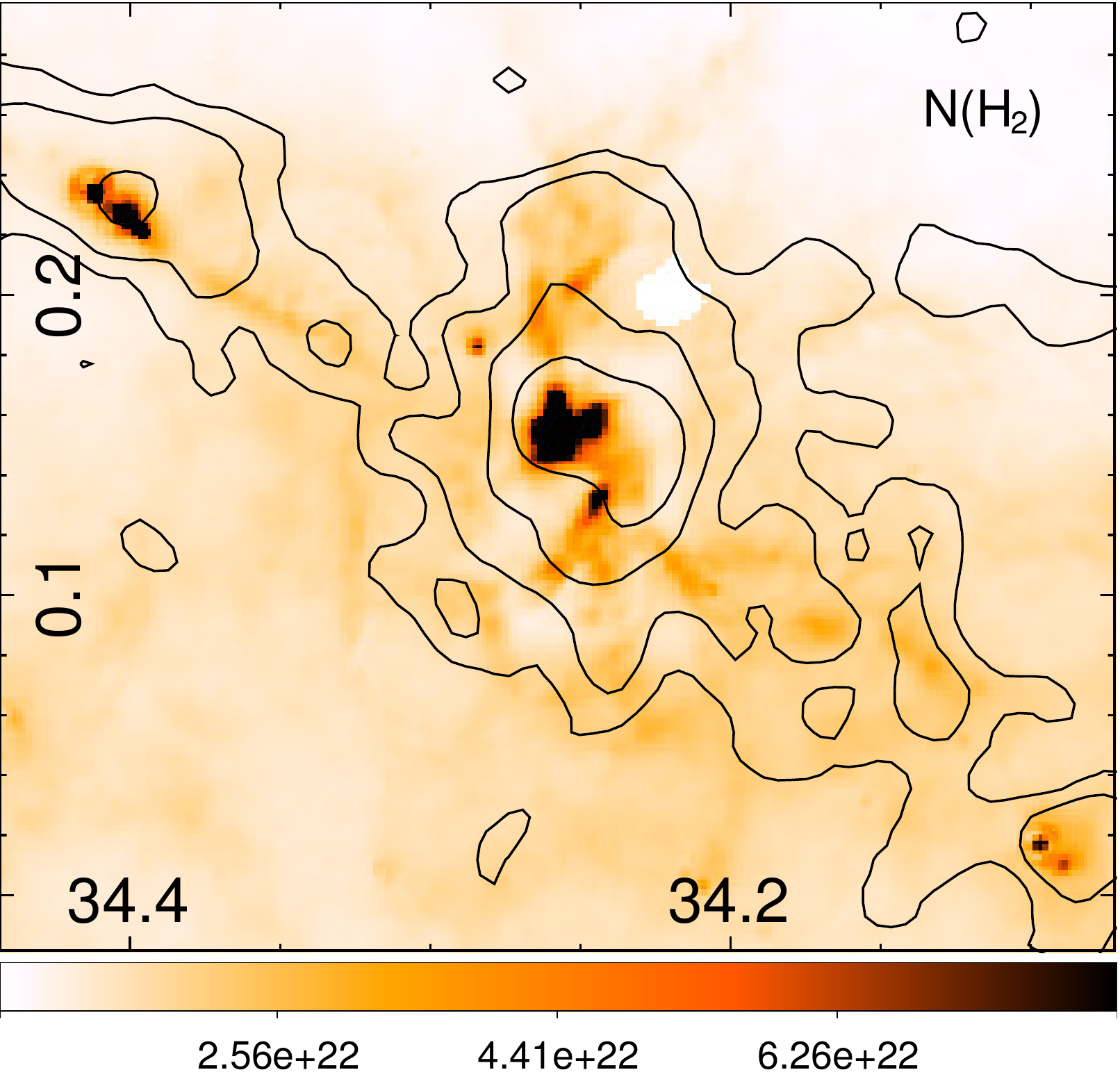}
\caption{Maps of dust temperature (top) and H$_{2}$ column density (bottom) obtained from http://www.astro.cardiff.ac.uk/research/ViaLactea/. The 
colourbars are in units of K and cm$^{-2}$, respectively. Contours of the integrated \8 J=1--0
are presented for reference.}
\label{sed}
\end{figure}

To make comparisons between the emissions of the molecular gas and the dust we obtain 
maps\footnote{http://www.astro.cardiff.ac.uk/research/ViaLactea/} of the dust temperature and H$_{2}$ column density 
N(H$_{2}$) (see Figure\,\ref{sed})
which were generated from the PPMAP procedure done to the Hi-GAL maps in the wavelength range 
70--500 $\mu$m \citep{marsh17}.

\begin{figure}[h!]
\centering
\includegraphics[width=8.1cm]{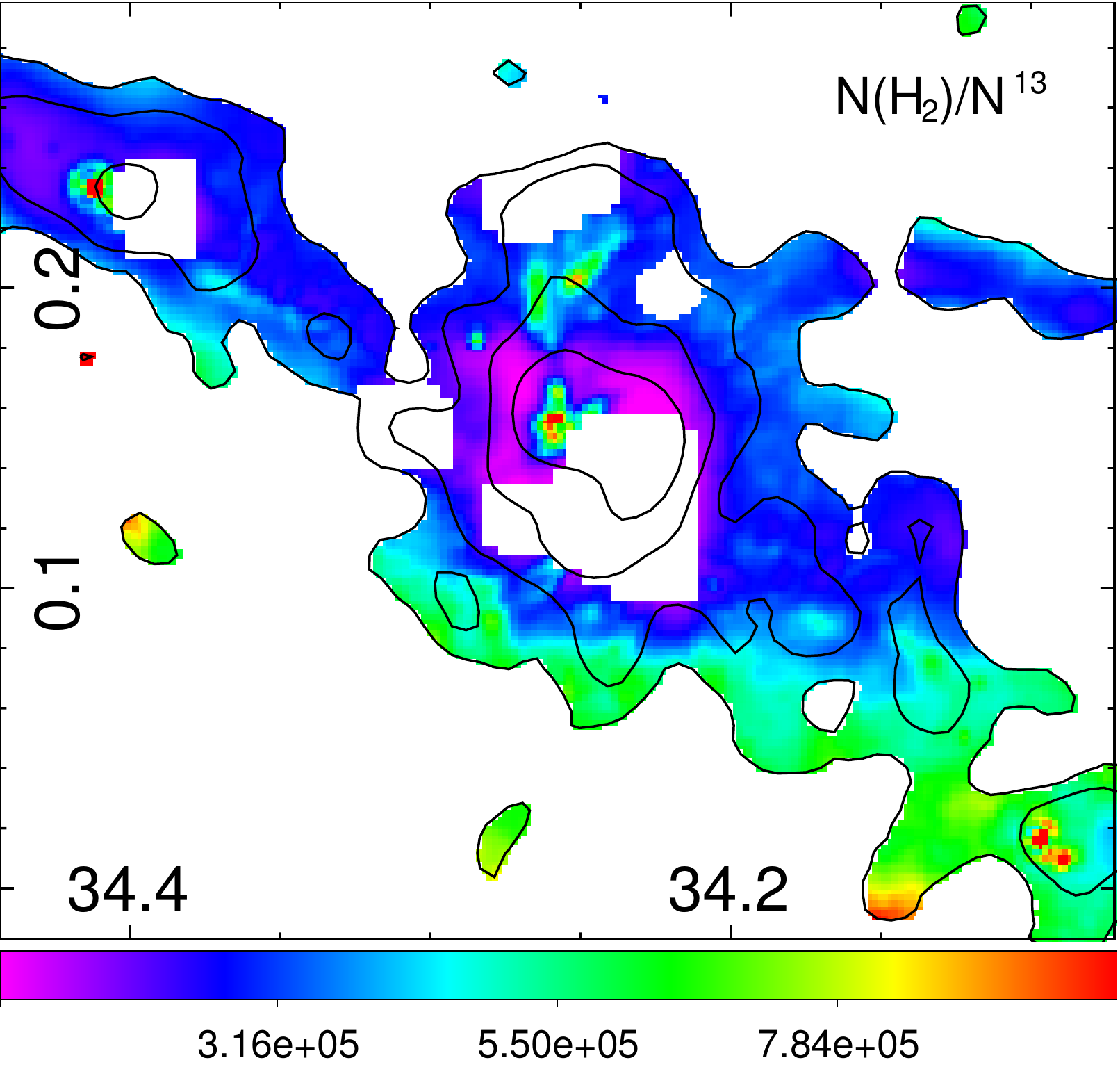}
\includegraphics[width=8.1cm]{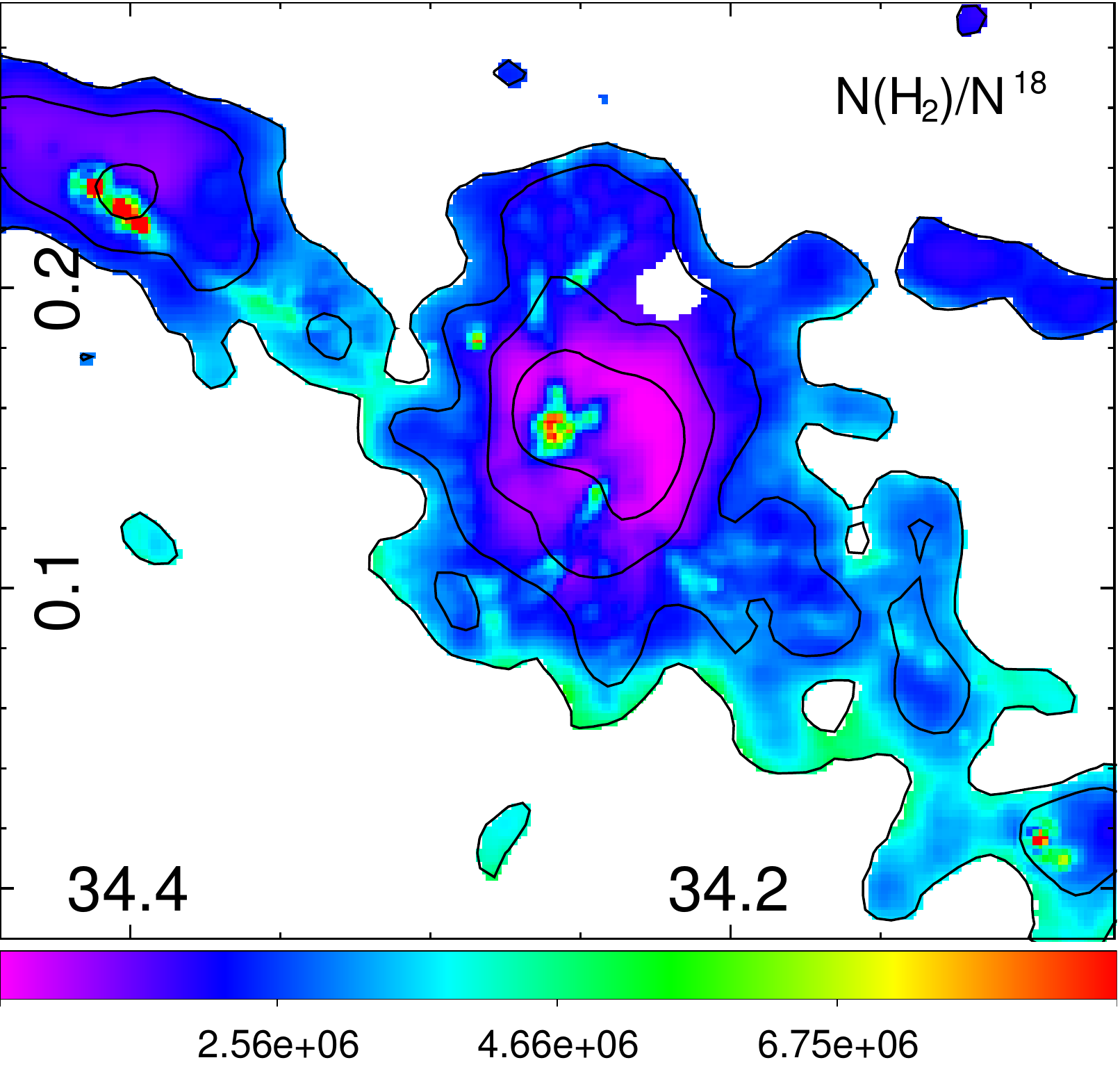}
\caption{Maps of N(H$_{2}$)/N(\3) and N(H$_{2}$)/N(\8). 
The angular resolution of these maps is about 36\s. Contours of the integrated \8 J=1--0 are presented for reference. }
\label{ratioH2}
\end{figure}

Ranges of dust temperature (T$_{dust}$) and N(H$_{2}$) with the average values are presented in Table\,\ref{sedr}. Additionally,
after convolving the N(\3) and N(\8) maps to the angular resolution of the N(H$_{2}$) map (about 36\s) we generate maps 
of N(H$_{2}$)/N(\3) and N(H$_{2}$)/N(\8) (see Fig.\,\ref{ratioH2}) whose
ranges and average values are also presented in Table\,\ref{sedr}.

\begin{table*}
\caption{Parameters derived from the IR data.}
\label{sedr}
\centering
\begin{tabular}{l c c c c}
\hline\hline
             & T$_{dust}$ &  N(H$_{2}$)                   &  N(H$_{2}$)/N(\3)    &  N(H$_{2}$)/N(\8)    \\
             &    [K]     &   [$\times10^{22}$ cm$^{-2}$] &    [$\times10^{5}$]  &   [$\times10^{6}$]                    \\
\hline
Ranges       &  18 -- 33   & 0.5 -- 3.8                    &  0.6 -- 7.0     &   0.4 -- 20     \\
Average      &   20.4      &    1.5                        &   3.9           &      3.0         \\
\hline
\end{tabular}
\end{table*}

\begin{figure}[h!]
\centering
\includegraphics[width=8.5cm]{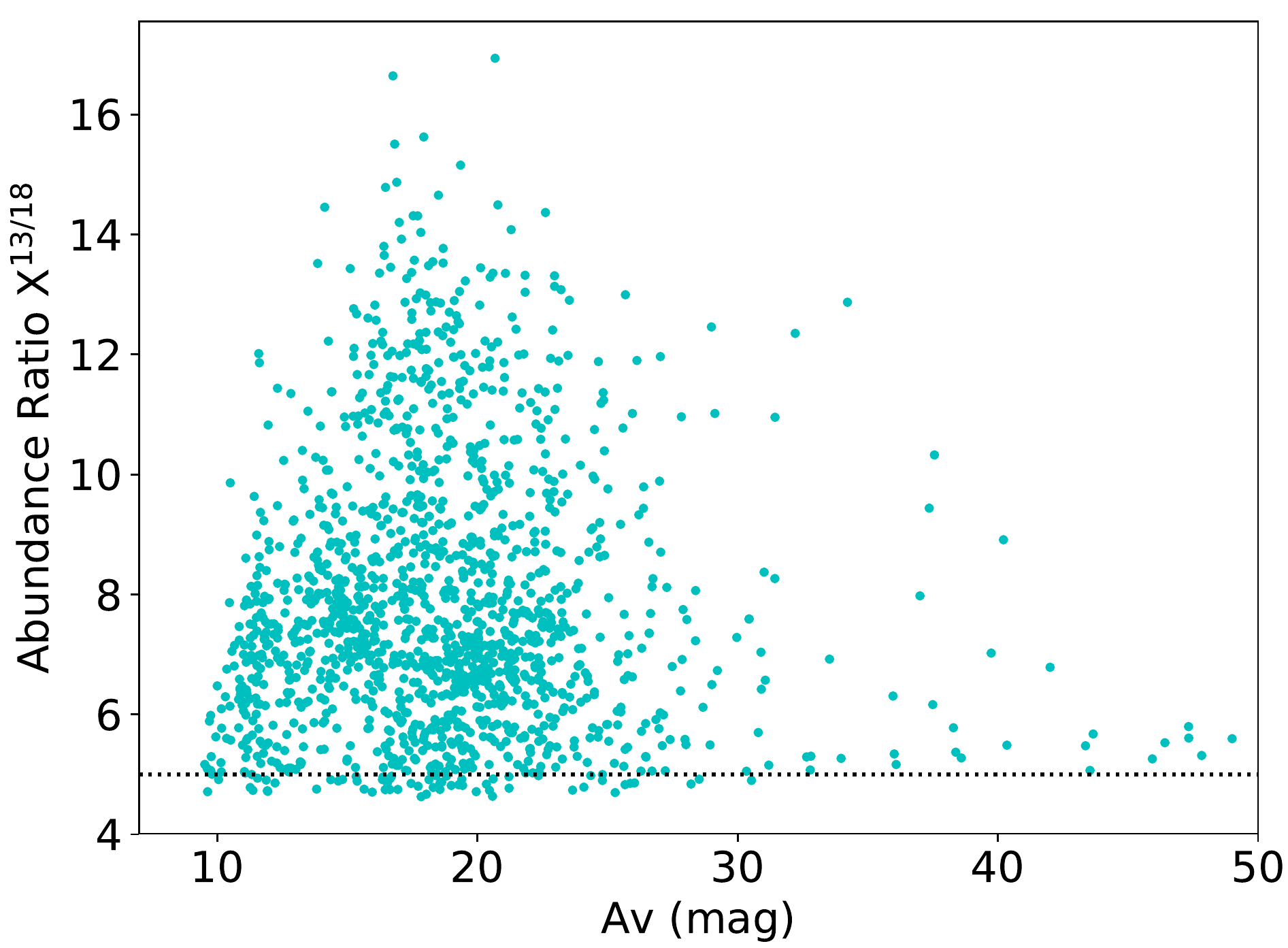}
\includegraphics[width=8.5cm]{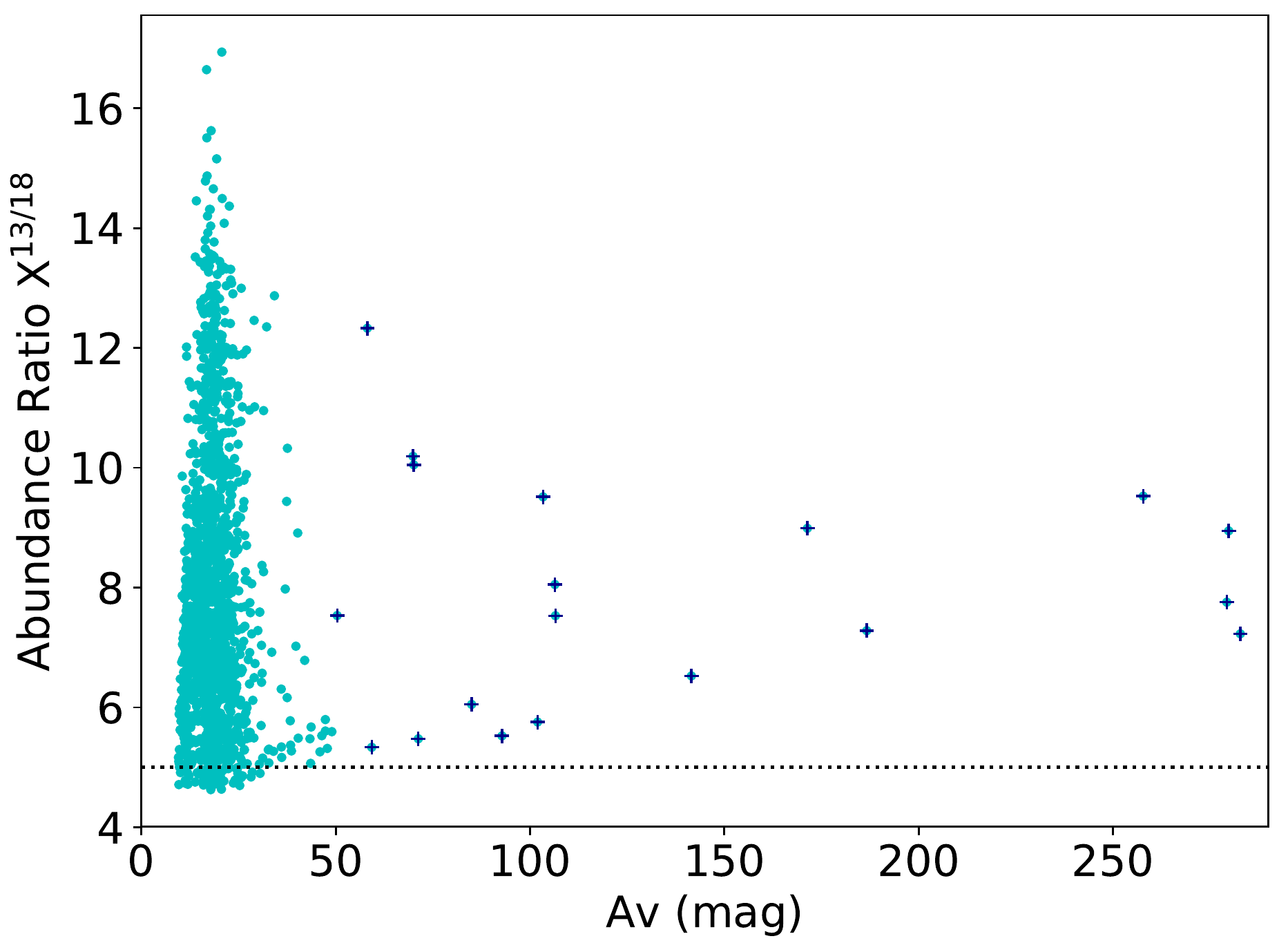}
\caption{Correlation between \x and the visual absorption A$_{\rm V}$ up to A$_{\rm V} = 50$ mag (top panel) and along the whole
A$_{\rm V}$ range (bottom panel). In bottom panel, the points corresponding
to pixels at the positions of the UC HII regions and their close surroundings are marked with blue crosses.}
\label{Avx}
\end{figure}

To investigate the dependence of \x~on the visual extinction towards this IRDC we obtain an A$_{\rm V}$ map from the 
N(H$_{2}$) following the relation from \citet{bohlin78}
\begin{equation}
A_{\rm V} = \frac{\rm N(H_{2})}{9.4 \times 10^{20} {\rm cm^{-2}}}.
\label{eqAv}
\end{equation}
The map of \x~was convolved to the angular resolution of the A$_{\rm V}$ map and in Figure\,\ref{Avx} we present plots
of \x~vs. A$_{\rm V}$.

\section{Discussion}

The molecular emission extends along the whole filamentary structure of IRDC 34.43$+$0.24, and as shown in Fig.\,\ref{integs}, 
the morphology of the \8 
emission is narrower and fits better with the infrared emission from IRDC than the \3 one.
The ranges of N(\3), N(\8) and \x~are quite similar to those found towards the Orion-A giant molecular
cloud by \citet{shima14}.
Additionally, our range of values of \x~also is in agreement with those values found towards LDN 1551, a nearby and isolated
star-forming region \citep{lin16}. 
Thus, our results show that \x~may have a similar behaviour among regions that are located at very different distances from us.
The obtained \x~average value (about 8) across the IRDC is larger than the solar system value of 5.5.

Assuming a distance of 3.9 kpc to IRDC 34.43$+$0.24, a galactocentric distance of about 5.7 kpc is obtained,
and following the relation between atomic ratios and the galactocentric distance presented in \citet{wilson94}, a \x of 7.4 is derived. 
While this value is in quite agreement with our 
average value of \x, it does not reflect the wide range of values found across the molecular cloud. Therefore, it is important to remark that 
assuming the `canonical' \x~from \citet{wilson94} may introduce some bias in the analysis of the molecular gas, mainly in regions 
that are exposed to FUV radiation.

Figure\,\ref{result10} (bottom panel) shows that highest values in \x~are found 
towards regions related to the interiors and/or borders of N61 and N62 bubbles. This 
can be explained through selective photodissociation of the \8 molecule due to the radiation responsible
of the bubbles generation, which may represents an observational support to the \8 selectively photodissociation
phenomenon, as found towards other nearby galactic regions (e.g. \citealt{yama19,lin16,kong15}) but in a quite distant filamentary IRDC. 
In general, an increase of \x~is observed near the PDRs (see the 8 $\mu$m emission in Fig.\,\ref{g34present}), precisely the regions in which
FUV photons are interacting with the molecular gas. From the FIR intensity we estimate the FUV radiation field, which
mainly reflects the action of the FUV photons in the dust. An increase of the FUV radiation field is observed towards the G34 complex,
where the UC HII region G34.26+0.15C lies, and towards the position of UC HII region G34.4+0.23 at the northeast of the IRDC 
(see Fig.\,\ref{uvfield}). By comparing the FUV map with the \x~map, two possible aspects of the influence of the UC HII regions in the abundance
ratio can be inferred: in clumps far from the UC HII regions, some \x~minimums coincide with \8~maximums, while in regions related to the clumps 
where the UC HII regions are embedded, this correlation is broken.
It is likely that the dust (which reflects the FUV field) may shield the molecular gas, and hence it is not possible to analyze a direct
correlation between the \x~factor and the FUV radiation field derived from the FIR emission. 

From a comparison between the radio continuum emission and the FUV radiation field (Figure\,\ref{uvfieldS}) it is worth noting that
the observed radio continuum shell, which represents ionized gas due to Lyman photons, is surrounded by the FUV radiation field, showing the
typical stratification of a PDR in HII regions (e.g. \citealt{draine11}).

By comparing the maps of \x and \R, i.e. comparing values that were derived from the LTE assumption with
values from direct measurements, a quite similar behaviour can be appreciated across the region.
The plot \x~vs. \R~shows a linear tendency with an increase in the slope as we analyze more dense regions in the cloud 
(see Fig.\,\ref{RvsX}). This behaviour may be explained by a major increase of $\tau^{13}$ with respect to $\tau^{18}$. On the other side, Region D 
shows a low correlation factor, which may be due to that in this region some \3 spectra present kinematic signatures of infall, 
and thus, the LTE assumption would no longer be valid in some pixels corresponding to it.

Assuming that the gas and dust are coupled, we can compare the \3 and \8 column densities obtained from the molecular lines with the N(H$_{2}$)
derived from the dust emission. The obtained average ratio N(H$_{2}$)/N(\3) of about $4\times10^{5}$ is in close agreement with the 
typical abundance ratio extensively used in the literature (N(H$_{2}$)/N(\3)$=5\times10^{5}$, e.g. \citealt{pineda08},
and references therein). The average ratio N(H$_{2}$)/N(\8) of $3.0\times10^{6}$ obtained in this work is in quite agreement with
the typical [H$_{2}$]/[\8] ratio ($5.8\times10^{6}$; e.g. \citealt{frer82}).

Additionally, we compare \x~with the visual absorption A$_{\rm V}$ derived from N(H$_{2}$). The maximum in \x~occurs between 
$17-22$ mag, and the behaviour of this relation is similar as found in previous works 
(e.g. \citealt{kong15,kim06,lada94}). In our case, an increase in \x~with respect A$_{\rm V}$ can be inferred towards 
the highest A$_{\rm V}$ values. We conclude that this plot shows selective photodissociation generated by both, 
external radiation (probably from the interstellar radiation field \citep{lin16}, and/or radiation from surrounding sources such as bubbles N61 and N62), 
and radiation from deeply embedded sources such as the UC HII regions.

\subsection{Beam filling factor effects and clumpiness}

It is known that the clouds are highly structured on the subparsec scale implying beam filling factors less than the unity.
We investigate the influence of a possible beam dilution effect in the \x.
The beam filling factor can be estimated from \citep{kim06}: 
$\phi=\frac{\theta^2_{\rm source}}{\theta^2_{\rm source}+\theta^2_{\rm beam}}$
where $\theta^2_{\rm source}$ and $\theta^2_{\rm beam}$ are the source and beam sizes, respectively.
The beam size of \3 and \8 J=1--0 data is 20\s, which corresponds to 0.37~pc at the 
distance of 3.9~kpc. From the emission with the best angular resolution used in this work, i.e. {\it Herschel}-PACS 70~$\mu$m 
(angular resolution $\sim$5\s), we determine that the sizes of clumps in the region are expected to exceed 0.7 pc. 
Thus, the beam filling factor of the used molecular data
is expected to exceed 0.4. Given that the \3 traces more extended molecular components than the \8,
we consider the limit case in which $\phi_{^{13}{\rm CO}}=1$ and $\phi_{\rm C^{18}O}=0.4$. Thus, \x~could be
overestimated up to 60\%. 

Taking into account that the beam dilution effect would be more important towards the densest/clumpy regions (mainly traced
by the \8 emission) than in regions with diffuse gas, the actual \x~value should be smaller than the measured one in clumpy regions. 
This is in agreement with \citet{zie00}, who point out that as the \8 emitting regions have much smaller extent than the \3 ones, resulting in a lower area filling factor, and also lower temperatures, larger (i.e. overestimated) \x~values may be observed.
Anyway, if the beam dilution is considered in these regions, the behaviour of \x~across the whole IRDC would highligth even more the effects of the 
selective photodissociation.

\section{Conclusions}

We carried out a large scale analysis of the abundance ratio \x~along the filamentary IRDC 34.43$+$0.24 using the \2, \3, and \3 J=1--0 emission with an angular resolution of about 20\s. 

In the line of previous works towards relative nearby molecular clouds, we find strong observational evidences supporting the \8 selectively photodissociation due to the FUV photons in a quite distance filamentary IRDC. 
A range of \x~between 3 and 30 was found across the molecular cloud related to the IRDC, with an average value of 8, larger than the typical solar value of about 5. 
From the FIR intensity we estimate the FUV radiation field, which mainly reflects the action of the FUV photons in the dust. 
We conclude that it is not possible to analyze a direct correlation between the \x~and the FUV radiation field derived from the FIR emission because it is likely 
that in some regions the dust may shield the molecular gas from the FUV photons.

From a comparison between \x~and the visual absorption A$_{\rm V}$, we conclude that selective photodissociation can be generated by both, the interstellar radiation field and/or radiation from surrounding HII regions (i.e. external radiation), and radiation from deeply embedded sources in the IRDC.

The average values of \x, N(H$_{2}$)/N(\3), and N(H$_{2}$)/N(\8), in which the N(H$_{2}$) was independently estimated from the dust emission, are in quite agreement with the `canonical' values used in the literature. However, works like this, show that if an accurate analysis of the molecular gas is required, the use of these `canonical' values may introduce some bias. In those cases, how the gas is irradiated by FUV photons and the implications of beam dilution should be considered, which can affect in different ways the abundance ratios.

\section*{Acknowledgements}

We thank the anonymous referee for her/his very helpful comments and suggestions.
M.B.A. and L.D. are doctoral fellows of CONICET, Argentina.
S.P. and  M.O. are members of the {\sl Carrera del Investigador Cient\'\i fico} of CONICET, Argentina.
This work was partially supported by Argentina grants awarded by UBA (UBACyT) and ANPCYT.
Nobeyama Radio Observatory is a branch of the National Astronomical Observatory of Japan, National Institutes of
Natural Sciences.

\bibliographystyle{pasa-mnras}
\bibliography{ref}

\end{document}